\documentclass[12pt,preprint]{aastex}
\usepackage{natbib}
\usepackage{graphicx}
\shorttitle{Existence of sterile neutrinos}
\begin{document}
\title{Observational evidences for the existence of 17.4 keV decaying 
degenerate sterile neutrinos near the Galactic Center} 
\author{M.~H.~Chan and M.~-C.~Chu}
\affil{Department of Physics and Institute of Theoretical Physics, 
\\ The Chinese University of Hong Kong, 
\\  Shatin, New Territories, Hong Kong, China}
\email{mhchan@phy.cuhk.edu.hk, mcchu@phy.cuhk.edu.hk}
\begin{abstract}
We show that the existence of a degenerate halo of sterile neutrinos with 
rest mass of 17.4 keV near the 
Galactic Center can account for both the excess 8.7 keV emission 
observed by the $Suzaku$ mission and the power needed ($10^{40}$ 
erg s$^{-1}$) to maintain 
the high temperature of the hot gas (8 keV) near the Galactic Center. The 
required decay rate and mixing angle of the sterile neutrinos 
are $\Gamma \ge 5 \times 10^{-20}$ s$^{-1}$ and $\sin^22 \theta \sim
10^{-3}-10^{-4}$ respectively. These values are consistent with a low 
reheating temperature in the inflation model, and suggest the 
exciting possibility that the sterile - 
active neutrino oscillation can be visible in near future experiments. 

\end{abstract} 

\section{Introduction}
Recent results from $Chandra$ indicate that soft ($\sim 0.8$ keV) and 
hard ($\sim 8$ keV) hot gas components exist within the inner 20 pc of the 
Galactic Center (GC) \citep{Park,Muno}. The power needed to 
maintain the 
temperatures of the soft and hard components of the hot gas are $3 \times 
10^{36}$ erg s$^{-1}$ and $10^{40}$ erg s$^{-1}$ respectively. The 
energy needed for the soft component can be explained by 1\% of kinetic 
energy from one supernova occurring every 3000 years, which is reasonable 
in the GC \citep{Muno}. However, the energy needed for the hard component 
cannot be explained satisfactorily \citep{Muno}. \citet{Chan} proposed 
that 
a decaying sterile neutrino halo existing near the GC can solve the 
problem. 
The photons emitted by the decays of the sterile neutrinos can 
heat up the surrounding 
gas, and the energy is subsequently transferred to the entire region at 
the GC. For this scenario to account for the observational data, the 
sterile neutrino rest mass is required to be $m_s \approx 16-19$ keV 
\citep{Chan}. Recently, $Suzaku$ X-ray mission has started to observe 
emission lines 
above 6 keV near the GC \citep{Koyama,Nobukawa}. The observed intensities 
of the emission lines, including 
Ly$\alpha$ (7.0 keV), Ly$\beta$ (8.2 keV) and Ly$\gamma$ (8.7 keV), from 
Fe XXVI are $1.66^{+0.09}_{-0.11} \times 10^{-4}$ ph cm$^{-2}$ s$^{-1}$, 
$2.29^{+1.35}_{-1.31} \times 10^{-5}$ ph cm$^{-2}$ s$^{-1}$ and 
$1.77^{+0.62}_{-0.56} \times 10^{-5}$ ph cm$^{-2}$ s$^{-1}$ respectively 
\citep{Koyama}. Based on these results, \citet{Prokhorov} find an excess 
of Ly$\gamma$ intensity of $(1.1 \pm 0.6) \times 10^{-5}$ ph cm$^{-2}$ 
s$^{-1}$, which cannot be explained by ionization and recombination 
processes. \citet{Prokhorov} proposed that decaying 
sterile neutrinos can provide the excess 8.7 keV photons. As the emitted 
photon energy $E_s \approx m_s/2$, the required $m_s$ is about 17 keV, 
which agrees with Chan and Chu's prediction (2008). 

By using the observed excess intensity of 8.7 keV photons and the 
Navarro-Frenk-White (NFW) density 
profile with 21.5 kpc scaled length to model the sterile neutrino halo, 
which means that the sterile 
neutrinos are the dominant dark matter component, \citet{Prokhorov} 
calculated the sterile neutrino decay rate and mixing angle with active 
neutrinos to be $\Gamma=(9.0 \pm 4.8) \times 10^{-28}$ s$^{-1}$ and  
$\sin^22 \theta=(4.1 \pm 2.2) \times 10^{-12}$ respectively. However, 
there is no 
evidence that the density profile of the sterile neutrino halo behaves 
like the NFW 
profile. If the sterile neutrinos are degenerate, 
the size of the halo can be very small (radius $R_s<1$ pc) 
\citep{Bilic,Chan2}. 
In this article, we extend our earlier work to account for the high 
temperature of the hot gas near the GC by including into the consideration 
of the $Suzaku$ excess of 8.7 keV photons; we show that the existence of  
a degenerate 
halo of decaying sterile neutrinos can account for both 
simultaneously \citep{Chan}. In this model, the 
optically dense 
gas clouds inside and nearby the sterile neutrino halo absorbs most of 
the energy of the 
photons emitted by the decays of the sterile neutrinos. The energy is then 
transferred to the surrounding gas 
clouds by conduction (mean free path $\sim 2$ pc). Since the optical 
depth is larger than 1, only a small portion of the decayed 
photons can escape from the GC and constitute the excess 8.7 keV 
emission. The calculated emission intensity from the 
decaying sterile neutrinos agrees with the observation as well as that 
required by 
\citet{Prokhorov}. In this scenario, the sterile neutrino decay rate 
$\Gamma \ge 5 \times 10^{-20}$ s$^{-1}$ 
and mixing angle $\sin^22 \theta \sim 10^{-3}-10^{-4}$, which are 
consistent with a low reheating 
temperature and suggest that sterile - active neutrino oscillation may be 
visible in near future experiments.

\section{The degenerate sterile neutrino halo model}
Sterile neutrinos may decay into active neutrinos and photons ($\nu_s 
\rightarrow \nu_a + \gamma$). The energy of the photons is assumed to be 
$E_s=8.7$ keV. Therefore, $m_s \approx 2E_s=17.4$ keV. Since the size 
of a degenerate sterile neutrino halo with total mass $M_s \le 
10^6M_{\odot}$ and $m_s \approx 17$ keV is much 
smaller than 20 pc, the total energy 
flux of the decayed photons within the field of view of $Suzaku$ (solid 
angle $\Omega$) is given by
\begin{equation}
F_s= \int_{\Omega} \frac{P}{4 \pi r_0^2}d \Omega \approx 2 \times 
10^{-11}~{\rm erg~cm^{-2}~s^{-1}},
\end{equation}
where $r_0=8.5$ kpc is the distance to the GC and $P=M_s \Gamma 
c^2/2=10^{40}$ erg s$^{-1}$ 
is the total power emitted in the sterile neutrino decays (the required 
power of the hard component of the hot gas near GC). The excess energy 
flux observed by $Suzaku$ is 
$F_s'=(1.1 
\pm 0.6) \times 10^{-5}E_s \approx (1.5 \pm 0.8) \times 10^{-13}$ erg 
cm$^{-2}$ s$^{-1}$. Therefore, only around 1\% of photons can escape 
from the GC, and most 
of the emitted photons are first absorbed by the gas clouds inside the 
sterile neutrino halo and nearby. The optical depth $\tau$ is given by 
\begin{equation}
\tau= \int \sum_in_i \sigma_i dr,
\end{equation}
where $\sigma_i$ is the 
effective absorption cross section of 8.7 keV photons by different gas 
components including cold molecular hydrogen gas ($i=$H$_2$), warm atomic 
gas 
($i$=H, He) hot ionized gas ($i$=hot) and very hot ionized gas ($i$=vhot),  
$n_i$ is the number density 
of the gas components. Assuming all the gas components follow the same 
density profile near the GC, which can be modelled by \citep{Schodel}
\begin{equation}
n_i= \frac{n_{0,i}}{(1+r/r_c)^{1.8}},
\end{equation}
where $r_c=0.34$ pc and $n_{0,i}$ are the core radius and core number 
density for the $i^{\rm th}$-type gas component. For H molecular and 
atomic gases, the 
average 
number densities within 450 pc are 74 cm$^{-3}$ and 0.9 cm$^{-3}$ 
respectively \citep{Ferriere}, which 
corresponds to $n_{0,H_2}=1.2 
\times 10^7$ cm$^{-3}$ and $n_{0,H}=1.5 \times 10^5$ cm$^{-3}$. Since 
hydrogen and helium constitute 70.4\% and 28.1\% by mass in the 
interstellar medium respectively 
\citep{Ferriere2}, we have $n_{0,He} \approx 0.1 \times 
(2n_{0,H2}+n_{0,H})=2.5 \times 10^6$ cm$^{-3}$. The hot ionic 
($10^5-10^6$ K) and very hot ionic ($10^7-10^8$ K) gas components 
constitute 4\%$-$5\% and 0.3\%$-$0.5\% by mass of all interstellar gas
respectively \citep{Ferriere}, 
which correspond to $n_{0, \rm hot}=(1-1.2) \times 10^6$ cm$^{-3}$ 
and $n_{0, \rm vhot}=(7-12) \times 10^4$ 
cm$^{-3}$. The 
effective photoionization cross sections of H$_2$, H, He, hot metal ions 
and very hot metal ions 
by 8.7 keV photons are $\sigma_{\rm H_2}=2.2 \times 10^{-26}$ cm$^2$, $ 
\sigma_{\rm H}=1.1 \times 10^{-26}$ 
cm$^2$, $\sigma_{\rm He}=3.8 \times 10^{-25}$ cm$^2$, $\sigma_{\rm 
hot}=(1.6-1.9) \times 10^{-24}$ cm$^2$ and $\sigma_{\rm 
vhot}=(1.5-1.6) \times 10^{-24}$ cm$^2$ 
respectively \citep{Yan,Chan,Daltabuit} \footnote{The effective 
cross sections of hot gas and very hot gas depend on the metallicity of 
the hot gas. The metallicities of 
Si, S and Fe in the interstellar medium are 1.13, 2.06 and 0.71 of solar 
metallicity respectively \citep{Muno}. 
The metallicity of other metal ions is assumed to be 2-3 
of solar metallicity \citep{Sakano}.}. 
Since the size of a degenerate sterile neutrino halo is very small, we 
can treat it as a point source. By using Eq.~(2) and the above data, we 
have
\begin{equation}
\tau \approx 1.25 \sum_i n_i \sigma_i r_c \approx 4.3 \pm 0.6.
\end{equation} 
Therefore, the observed 8.7 keV photon flux should be
\begin{equation}
F_s'=F_se^{-\tau}=(3.2 \pm 1.7) \times 10^{-13}~\rm erg~cm^{-2}~s^{-1},
\end{equation}
which is consistent with the observational data $F_s'=(1.5 \pm 0.8) \times 
10^{-13}$ erg cm$^{-2}$ s$^{-1}$. 

Observational data on stars 
S1 and S2 near the GC constrain $M_s$ to be less than or equal to $2 
\times 10^5M_{\odot}$ 
\citep{Schodel}. Since in our model the sterile neutrino decays also 
supply the 
energy needed for the hard component of the hot gas near the GC, $M_s 
\Gamma c^2/2=10^{40}$ erg s$^{-1}$, we have $\Gamma \ge 5 \times 
10^{-20}$ s$^{-1}$, which coincides with what is needed to solve the 
cooling flow problem in galaxy clusters \citep{Chan2} \footnote{Since the 
sterile neutrinos can 
also decay into 3 active neutrinos, if the radiative decay rate 
$\Gamma \sim 10^{-19}$ s$^{-1}$, the total decay rate $\approx 128 \Gamma 
\sim 10^{-17}$ $s^{-1}$ \citep{Barger}, which matches with
\citet{Chan2}'s prediction.}. The mixing angle of the 
sterile neutrinos is given by \citep{Barger}
\begin{equation}
\sin^22 \theta= 1 \times 10^{-3} \left( \frac{\Gamma}{2 \times 
10^{-19}~\rm s^{-1}} \right) \left( \frac{m_s}{17.4~\rm keV} \right)^{-5}.
\end{equation}
In our model, the mixing angle is $\sin^22 \theta \sim 10^{-3}-10^{-4}$, 
which seems to disagree with the standard non-resonant 
production mechanism \citep{Dodelson}. Nevertheless, in the low reheating 
temperature scenario, the number density of active 
neutrinos is lower and the mixing angle can be much larger than the 
standard prediction \citep{Giudice}. \citet{Gelmini} proposed that if the 
reheating temperature is lower than 5 MeV, the mixing angle can 
be as 
large as $\sin^22 \theta \sim 10^{-3}$, which is consistent with our 
results. With such a large mixing angle, the sterile - active neutrino 
oscillation may be visible in future experiments \citep{Gelmini,Yaguna}. 

\section{Summary}
We have shown that the excess 8.7 keV emission observed by the {\it 
Suzaku} X-ray mission as well as the high temperature ($\sim 8$ keV) of 
the hot gas near the GC can both be accounted for by the 
existence of a halo of degenerate decaying sterile neutrinos with 17.4 keV 
rest mass. The emitted photons from the sterile neutrino halo hiding 
deeply at the GC heat up the surrounding gas. The energy 
is then transferred to the nearby gas clouds to maintain their temperature 
at around 8 keV. The calculated photon flux for 8.7 keV photons from the 
GC is $F_s'=(3.2 \pm 1.7) \times 10^{-13}$ erg cm$^{-2}$ s$^{-1}$, which 
is consistent with the observed flux $F_s'=(1.5 \pm 0.8) \times 
10^{-13}$ erg cm$^{-2}$ s$^{-1}$. 

In this scenario, the decay 
rate and mixing angle of the sterile neutrinos are given by $\Gamma \ge 
5 \times 10^{-20}$ s$^{-1}$ and 
$\sin^22 \theta \sim 10^{-3}-10^{-4}$ respectively. 
These values are also consistent with those needed to account for the 
cooling flow problem in galaxy clusters 
\citep{Chan2} and are consistent with the low reheating temperature model 
\citep{Gelmini,Gelmini2}. The relatively large mixing angle suggests that 
sterile neutrinos may be directly studied in lab experiments in the near 
future, making it a particularly exciting candidate of warm dark matter 
particles. 

\section{acknowledgements}
This work is partially supported by grants from the Research Grant 
Council of the Hong Kong Special Administrative Region, China (Project 
Nos. 400805 and 400910).

\end{document}